\documentclass[11pt]{article}    
\usepackage{emulateapj}	  
\usepackage{flushrt,natbib}     
\usepackage{epsfig,journals}	 
\newcommand{\Bcoeff}{\Lambda}
\newcommand{\rhob}{\rho_0} 
\newcommand{\Ib}{{\cal I}} 
\newcommand{\Sb}{{\cal S}}
\newcommand{\Stot}{{\cal S}_{\rm tot}}
\newcommand{\IbOne}{{\cal I}_1} 
\newcommand{\Enhc}{{\cal E}} 
\newcommand{\mb}{{m}} 
\newcommand{\rb}{{\ell}} 
\newcommand{\rmOne}{{r_{m1}}} 
\newcommand{\tb}{{t}} 
 
\newcommand{\pic}{{\Pi_c}} 
\newcommand{\piacc}{{\Pi_{\rm acc}}} 
\newcommand{\SFRff}{{\mbox{SFR}_{\rm ff}}} 
\newcommand{\Mstar}{{M_\star}} 
\newcommand{\Mbarstar}{{\overline{M}_\star}} 
\newcommand{\vch}{{v_c}}

\newcommand{\convln}{\lower.5ex\hbox{$\; \buildrel \otimes \over \ln \;$}}
\newcommand{\ltsim}{\lower.5ex\hbox{$\; \buildrel < \over \sim \;$}}
\newcommand{\gt}{\lower.5ex\hbox{$\; \buildrel > \over \sim \;$}}
\newcommand{\Iesc}{{\cal I}_{\rm esc}}
\newcommand{\vesc}{v_{\rm esc}}
\begin{document}

\title{Protostellar outflow-driven turbulence}

\author{Christopher D. Matzner}
\affil{Department of Astronomy \& Astrophysics \\ University of
Toronto, 60 St.~George Street, Toronto, ON M5S 3H8, Canada}

\begin{abstract}
Protostellar outflows crisscross the regions of star cluster
formation, stirring turbulence and altering the evolution of the
forming cluster.  We model the stirring of turbulent motions by
protostellar outflows, building on an observation that the scaling
law of supersonic turbulence implies a momentum cascade analogous to
the energy cascade in Kolmogorov turbulence.  We then generalize this
model to account for a diversity of outflow strengths, and for outflow
collimation, both of which enhance turbulence.  For a single value of
its coupling coefficient the model is consistent with turbulence
simulations by Li \& Nakamura and, plausibly, with observations of the
NGC\,1333 cluster-forming region.  Outflow-driven turbulence is strong
enough to stall collapse in cluster-forming regions for several
crossing times, relieving the mismatch between star formation and
turbulent decay rates.  The predicted line-width-size scaling implies
radial density indices between -1 and -2 for regions supported by
outflow-driven turbulence, with a tendency for steeper profiles in
regions that are more massive or have higher column densities.
\end{abstract}

\section{Introduction}\label{intro}

Star formation in the Milky Way is known to occur predominantly within
star clusters.  Clusters with different multiplicites ($N_\star$) are
born in a remarkably flat distribution, $d\dot{\cal N}_{\rm cl}/d\ln
N_\star\propto N_\star^{-1}$. This is flat in the sense that the birth
cohort of a given star is equally likely to be in any decade of
$N_\star$ within the allowed range, from a lower limit of $N_\star
\sim 50-100$ to an upper limit that depends on the area sampled -- a
few thousand, based on surveys out to 2 kpc
\citep{2003ARA&A..41...57L}, or about half a million
\citep{1997ApJ...476..144M}, based on HII region surveys of the entire
inner Galaxy.  The number, density, and lifetime of the birth cohort
affect the likelihood of exposure to ionizing radiation and supernovae
from the massive cluster members, as well as the likelihood of close
stellar encounters, all of which affect the formation and evolution of
planetary systems.  

The Solar System, for instance, is thought to derive from a cluster
with a few thousand members \citep{2001Icar..150..151A} on the basis
that a nearby supernova supplied short-lived radionuclides, and that
passing stars did not disturb the outer planets.  The orbit of the
planetoid Sedna \citep{2005ApJ...635L..97B} appears to require an
encounter with another star well within a thousand AU, an event which is
nearly impossible except in the dense (but not too dense) context of
the Sun's birth cluster.

There is growing evidence that star cluster formation is a slow
process, in the sense that the star formation rate, though quite
uncertain, is about an order of magnitude slower than the mean
free-fall rate of the parent gas ``clump''.
\cite{1999ApJ...525..772P} and \cite{2006ApJ...644..355H} argue from
the Orion Nebula Cluster's Hertzsprung-Russell diagram that star
formation began there at least 10 Myr ago, as has been dramatically
confirmed by the detection of lithium depletion in some of its members
\citep{2005ApJ...626L..49P}.  The common origin of runaway stars
AE~Aurigae and $\mu$~Columbae in the cluster's vicinity 2.6 Myr ago
\citep{1954ApJ...119..625B} supports this assertion
\citep{2006ApJ...641L.121T}.  \citeauthor{2006ApJ...641L.121T} also
point to the fact that cluster mass profiles are quite smooth,
requiring time for subclusters to merge.  \cite{2006astro.ph..6277K}
compute the star formation rate parameter $\SFRff$, which is
normalized to the free-fall rate, from molecular and infrared
observations of dense, cluster-forming clumps, finding $\SFRff\sim
10^{-1.7}$ (with significant uncertainty) for hydrogen densities
ranging from $n_H = 10^2$ to $10^{5.3}$\,cm$^{-3}$.  As they point
out, this is consistent both with theories in which star formation is
regulated by ambipolar diffusion within magnetized gas \citep{M89},
and with those in which turbulence suppresses the rate of localized
collapse \citep{2005ApJ...630..250K}.  Regardless of its cause, slow
cluster formation is in sharp contrast with the rapid decay of
supersonic turbulence over at most a couple free-fall times
\citep[][]{1998ApJ...508L..99S, 1999ApJ...524..169M}.  Therefore,
forming star clusters must tap some source of energy to sustain the
observed turbulence -- and the nature of this feedback may leave its
imprint on the clusters' properties.

This paper addresses the driving of turbulence by protostellar jets
and winds, which are thought to emerge from stars of all masses as
they form.  Since outflows driven by these jets and winds are
ubiquitous, outflow-driven turbulence is a pervasive and inevitable
consequence of stellar cluster formation.

A second, more sporadic form of feedback comes from the most massive
cluster members, whose ionizing radiation is a strong driver of
turbulence within giant molecular clouds
\citep[GMCs;][]{2002ApJ...566..302M, 2006astro.ph..8471K}.  However,
cluster-forming clumps are much denser than the GMCs surrounding them
($\sim 10^{4-5}$ rather than $\sim 10^{2}$ H atoms cm$^{-3}$), and
high densities damp the dynamical effects of
photoionization.  We shall therefore concentrate on outflow-driven
turbulence.  (See \citealt{TanMcKee00} for preliminary cluster-formation
models which include photoionization.)

\cite{NS80} proposed that protostellar outflows sustain the turbulence
in GMCs, before it was known that star formation is restricted to the
densest and most massive clumps within them.  \cite{M89} adopted
protostellar outflows as the driving agents in his equilibrium theory
of GMCs.  Concentrating on the scale of a cluster-forming clump,
\cite{myphd} and \cite{1999sf99.proc..353M} incorporated outflows into
dynamical models for star cluster formation.  Recent simulations of
cluster formation by \cite{2006ApJ...640L.187L} provide an important
demonstration that protostellar outflows are dynamically important in
this process.

The theory of outflow-driven turbulence presented here improves upon
these previous studies by accounting for two of the outflows' key
properties: a strong collimation toward an outflow axis, and a
diversity in strength associated with the range of possible stellar
masses.  These effects conspire to {\em enhance} the outflows' effect
on the dynamics of cluster-forming clumps, relative to a model in
which outflows are spherical and all have the same strength.

In \S\,\ref{oom} we establish the basic dimensional scales and
dimensionless ratios that define outflow-driven turbulence.  In \S
\ref{Spectrum} we propose models for the energy spectrum and
line-width-size relation of outflow-driven turbulence, drawing on an
insight about the dynamical nature of supersonic turbulence in
strongly radiative gas.  We extend this model to a diversity of
outflow intensities in \S\,\ref{variations}, and evaluate it for the
stellar initial mass function (IMF) and outflow collimation in \S
\ref{IMF} and \ref{collimation}, respectively.  The loss of momentum
from a finite region due to collimated outflows is considered in
\S\,\ref{Nonlocality}.  We compare to available simulations in
\S\,\ref{sim} and to observations of NGC\,1333 in \S\,\ref{obs}, before
drawing conclusions about the dynamics of stellar cluster formation in
\S\,\ref{implic}.

\section{Dimensional analysis}\label{oom}

To begin as generally as possible, let us first consider a pressureless
medium of mean density $\rhob$ stirred by spherical
explosions whose characteristic momentum is $\Ib$, that occur at a rate per
unit volume $\Sb$.  We choose to describe these by their momentum rather
than their energy, because we expect the shocked material to be highly
radiative.   These three parameters with independent dimensions define 
scales of mass, length, and time: 
\begin{equation} \label{scales}
\mb ={ \rhob^{4/7} \Ib^{3/7} \over \Sb^{3/7}}, ~
\rb = { \Ib^{1/7} \over \rhob^{1/7}\Sb^{1/7}}, ~ 
\tb = {\rhob^{3/7} \over \Ib^{3/7} \Sb^{4/7}}. 
\end{equation}
One quantity of interest is the turbulent line-width, i.e., 
the characteristic turbulent velocity, measured on scale $\rb$: 
\begin{equation} \label{SigmaAtRb}
{\rb\over\tb} = {\Ib^{4/7} \Sb^{3/7}\over \rhob^{4/7}}. 
\end{equation}
The dynamical state of the gas will be highly turbulent and clumpy,
but in the absence of other dimensional parameters any quantity must
obey these scalings.  There is one explosion, on average, per $\rb^3
\tb$.  Intermittent explosions are the dominant feature of flow on
scales smaller than $\rb$, whereas flow on larger scales reflects many
overlapping explosions.

To estimate characteristic scales for the cluster-forming regions
discussed later, consider $\Ib= 10^{39.6}$\,g\,cm\,s$^{-1}$, $\Sb =
10^{-67.2}$\,cm$^{-3}$\,s$^{-1}$, and $\rhob =
10^{-19.6}$\,g\,cm$^{-3}$ -- values typical of cluster-forming regions
like NGC\,1333 (\S~\ref{obs}).  Then, $\rb = 0.38$\,pc,
$\tb=0.34$\,Myr, $\mb = 19\,M_\odot$, and $\rb/\tb =
1.1$\,km\,s$^{-1}$.  Outflow-driven turbulence (1) extends well beyond
the regions of localized collapse associated with individual forming
stars, and (2) involves motions significantly faster than the thermal
sound speed ($c_s\sim 0.2$\,km\,s$^{-1}$).  In \S\,\ref{variations} we
shall find that these qualities, which distinguish outflows as likely
agents of support, are only enhanced by effects like outflow
collimation.

\subsection{Gravity, stellar mass, and finite-size regions} \label{Grav}
We now introduce gravity, and restrict our attention to a region of
finite size $R$ -- hence of mass $M= 4\pi R^3\rhob/3$ if we take
$\rhob$ to be the mean density.  Gravity and finite radius come
together naturally, because star formation is known to occur in
self-gravitating ``clumps'' of finite mass and radius.

Let us also introduce a stellar mass scale $\Mstar$, and suppose that
the explosion momentum is related to $\Mstar$ through a characteristic
velocity:
\begin{equation} \label{I=Mstar_vch}
\Ib = \Mstar \vch. 
\end{equation} 
If we consider the stellar mass function fixed, we might take $\Mstar
\simeq 0.5 M_\odot$; \cite{1999ApJ...526L.109M} adopt $\vch\simeq
40$\,km\,s$^{-1}$ in protostellar outflows, implying $\Ib\simeq
10^{39.6}$\,g\,cm\,s$^{-1}$.

Given $\Mstar$, we may define
\begin{equation}\label{SFR}
\Sb = \left(32\over3\pi\right)^{1/2}\SFRff {G^{1/2}\rhob^{3/2}\over
\Mstar}
\end{equation} 
where $\SFRff$ is the star formation rate normalized to the free fall
rate at the mean density.  \cite{M89} estimates this parameter under
the assumption that ambipolar diffusion determines the rate of
collapse, and finds $\SFRff \simeq 0.08$ in regions well shielded from
external far-ultraviolet radiation.  The turbulence-regulated star
formation theory of \cite{2005ApJ...630..250K} predicts $\SFRff \simeq
0.034$ in nearby star-froming clumps.  

Although we used $\Mstar$ in defining $\vch$ and $\SFRff$, it is
important to realize that outflow-driven turbulence is invariant under
the transformation $M_\star\to\alpha M_\star$,
$\SFRff\to\alpha\SFRff$, and $\vch\to\alpha^{-1} \vch$, which leaves
$\Sb$ and $\Ib$ unchanged.  Therefore $\Mstar$ introduces no 
new dimensional scale so far as turbulence is concerned.  A corollary
is that, although $\vch$ enters explicitly into the star formation
efficiency \citep{2000ApJ...545..364M}, it will always appear in the
combination $\SFRff\vch\propto \Sb\Ib$ in dimensionless ratios given
here.

Self-gravity and finite radius, on the other hand, introduce the
dimensional quantities $G$ (Newton's constant) and $R$, which can be
used to form two dimensionless ratios with $\Sb$, $\Ib$, and $\rhob$.
Useful forms include the radius ratio 
\begin{equation}\label{PiR}
\Pi_R\equiv
{R\over\rb},
\end{equation} 
the acceleration ratio
\begin{equation} \label{PiAcc} 
\piacc \equiv {\rb \over \tb^2}{R\over v_K^2} =
{\Ib\Sb\over\rhob} {R\over v_K^2} = \left(2\over\pi\right)^{3/2}
\SFRff {\vch\over v_K}, 
\end{equation} 
which measures the acceleration scale $\Sb\Ib/\rhob$ in units of the
surface gravity $v_K^2/R = GM/R^2$ (for Kepler speed $v_K$), and the
quantity $\Iesc/\bar \Ib\simeq \Pi_{\rm acc}^{-1/2} \Pi_R^{7/2}$ which we
define in \S\,\ref{Nonlocality}.  

The scalings in equations (\ref{scales}) and (\ref{SigmaAtRb}) pertain
to an infinite ($\Pi_R\gg1$) and non-self-gravitating ($\piacc\gg1$)
medium rather than a realistic clump or cloud.  Consider the opposite:
if $\Pi_R\ll1$, then the region will be cleared by a typical outflow;
and if $\Pi_{\rm acc}\ll1$, then outflows cannot provide the dynamical
pressure required for cloud support.  One might therefore expect both
parameters to be $\gtrsim 1$ in a real cluster-forming environment.
More specific predictions are provided in \S\,\ref{Nonlocality} where
we detail a model for turbulence in a cloud of finite radius and mass,
and in \S\,\ref{implic} where we consider the cloud's dynamical
expansion or contraction in response to outflow-generated turbulence.

\subsection{Finite sound speed and magnetization}\label{thermal}
Continuing in the spirit of dimensional analysis, let us allow the gas
to have a finite sound speed $c_s$; the relevant dimensionless
parameter is 
\begin{equation} \label{PiC}
\pic = {c_s \tb\over \rb} 
\end{equation} 
which is an estimate for  the inverse of the turbulent Mach
number on scale $\rb$.  Turbulence is supersonic and strongly compressive
so long as $\pic \ll 1$; for the parameters adopted above, this is
true if $\SFRff \gg 10^{-3.2} n_{H4}^{-1/6}
(c_s/0.2{\rm\,km\,s^{-1}})^{7/3}$,  where the hydrogen density is $10^4 n_{H4}$
cm$^{-3}$.  So long as $\pic\ll1$ we expect outflow-driven turbulence
to be insensitive to the precise value of $c_s$; this is supported by
the more detailed calculations given below in \S\,\ref{variations}.

The influence of sound speed on the star formation process itself
should not be understated, however.  By setting the thermal Jeans
scale, finite sound speed determines both the normalization of stellar
masses \cite[e.g.,][]{2002ApJ...576..870P}, and the star formation
rate \citep[at least in the model of][]{2005ApJ...630..250K}, thereby
affecting $\Sb$ and $\Ib$.

The dynamics of the gas will certainly be affected by magnetic fields
\citep[e.g.,][]{PPIII}.  These introduce at least two
additional dimensionless parameters: the degree of magnetization, as
measured by the global mass-to-flux ratio or the local plasma-$\beta$
parameter; and the ambipolar diffusion rate, as measured by the ratio
of the neutral-ion collision time to the free-fall time, say.
We expect that, like $\pic$, these parameters influence the
coefficients that appear below -- the coupling factor $\Bcoeff$
defined in \S\,\ref{Spectrum}, and the critical value of $\alpha_{\rm
  vir}$ in \S\,\ref{implic} -- but introduce no other qualitative
change to the theory.  This expectation can only be tested by
comparing the theory to observations and simulations.

\section{Outflow momentum per stellar mass, $\vch$} \label{vch} 

The outflow momentum per unit stellar mass, $\vch$, was estimated to
be $40$\,km\,s$^{-1}$ by \cite{2000ApJ...545..364M} on the basis that
the typical protostellar wind velocity is $\sim$ 200\,km\,s$^{-1}$ and
about one-sixth of the accreted mass is ejected as wind -- a value
which is intermediate among several theoretical predictions.  In order
to check our estimate of $\vch$, we combine the
\cite{1992ApJ...392..667P} evolutionary models for accreting
protostars with an observational inference by \cite{Richeretal} of the
relation between stellar and outflow properties.  For those stars
whose luminosity is dominated by accretion, \citeauthor{Richeretal}
show that the outflow momentum is roughly $0.3 v_{k\star}$, where
$v_{k\star}$ is the Kepler velocity at the surface of the protostar.
We obtain $v_{k\star}$ from \citeauthor{1992ApJ...392..667P}'s models,
and apply the \citeauthor{Richeretal} rule to all the material
accreting onto a star of final mass $M_\star$ in order to calculate
its net outflow momentum $\Ib(M_\star)$.  Dividing the total momentum
by the total mass for a stellar population drawn from the
\cite{2001MNRAS.322..231K} IMF with an upper cutoff of $M_{\star u}$, we find 
\begin{equation} \label{vch-Estimate}
\vch \simeq (28,27,22) \log_{10}{M_{\star u}\over0.11\,M_\odot}
\end{equation} 
assuming stars all accrete at $(10^{-5}, 3\times10^{-5},
10^{-4})\,M_\odot$\,yr$^{-1}$, respectively.  The result is lower for a higher
accretion rate, because these stars have less time to contract during
accretion.

For an upper cutoff $M_{\star u}=120\,M_\odot$ this exercise gives
about twice our fiducial estimate of 40\,$\rm km\,s^{-1}$, as
previously noted by \cite{2002ASPC..267..267T}.  However, the excess
momentum comes from massive stars that (1) require an extrapolation of
the \cite{Richeretal} scaling, (2) form only sporadically, relative to
low-mass objects, and (3) may not form at all, if the cluster in
question is too small to sample the IMF.  For simplicity we hold to
the initial estimate $\vch= 40$\,km\,s$^{-1}$ for the purpose of
making estimates, although a more complicated model could be
accommodated using the theory we present in \S~\ref{variations}.

\section{Velocity spectrum and line-width-size relation}
\label{Spectrum}  

The winds that drive outflows possess a high energy per unit mass,
which reflects the depth of the potential well from which they were
launched.  Most of this is lost to radiation at the wind deceleration
shock.  What remains is the kinetic energy of the expanding shell, but
this declines in proportion to the shell velocity as additional mass
is swept up.  At a radius of order $\rb$ the shell loses coherence and
merges with supersonic turbulent motions -- which themselves dissipate
energy, as described, for instance, by \cite{1998ApJ...508L..99S},
\cite{1998PhRvL..80.2754M}, and \cite{1999ApJ...524..169M}.  In
contrast to the Kolmogorov cascade of incompressible turbulence, this
supersonic cascade is not described by a constant energy flux to small
scales.

On the other hand, strongly radiative outflow shells conserve momentum
in each direction, and therefore conserve a net scalar momentum $\Ib$ as
they expand
-- where by ``scalar momentum'' we mean 
\[ \int \rho |{\mathbf u}-{\mathbf u}_{\rm cm}| d^3 {\mathbf x}\] 
if $\rho$ and $\mathbf u$ are the local density and fluid velocity,
${\mathbf u}_{\rm cm}$ is the center-of-mass velocity, and the
integral extends over the volume affected by the outflow. 
A population of expanding outflows delivers momentum at
a rate $\Sb\Ib$ per unit volume.  Turbulence is therefore driven, on
scale $\rb$, by a characteristic acceleration $\Sb\Ib/\rhob$.

Does a radiative, supersonic cascade transport {\em momentum} to small
scales the way an incompressive cascade transports energy?  This seems
plausible, based on the dynamics of expanding outflow shells.
Moreover, this hypothesis is consistent with the well-known scaling laws of
supersonic turbulence.  The one-dimensional (spherically averaged)
spectrum of the velocity field $\mathbf u(\mathbf x)$ is defined as
\begin{equation}\label{defineSpectrum}
E(k) = \frac{1}{2\pi}\int_{0}^{\infty}  \left<{\mathbf
      u}({\mathbf x}) 
\cdot {\mathbf u}({\mathbf x}+ {\mathbf n}\,r)\right> e^{-i k r} dr
\end{equation} 
where angle brackets indicate an average over time, space, and the
direction of the unit vector $\mathbf n$.  Suppose the turbulence
transports an acceleration $\Sb\Ib/\rhob$ from $\rb$ to small scales.
As $S$ has units cm$^{-3}$\,s$^{-1}$ and as $\Sb\Ib/\rhob$ has units
cm\,s$^{-2}$,
\begin{equation}\label{spectrum} 
E(k) = A {\Sb\Ib\over \rhob} k^{-2}~~~~(k\gg \rb^{-1})
\end{equation}
for some coefficient $A$ which depends on other dimensionless
parameters.  Equation (\ref{spectrum}) matches the spectral slope for
supersonic turbulence, as determined by numerical experiment
\citep{1992PhRvL..68.3156P}.  
It coincides with the spectral slope in Burgers' turbulence
\citep{1979JFM....93..337K}, and indeed with that of individual
shocks.  Moreover it resembles \citeauthor{1981MNRAS.194..809L}'s
(\citeyear{1981MNRAS.194..809L}) and matches
\citeauthor{1987ApJ...319..730S}'s (\citeyear{1987ApJ...319..730S})
estimates of the line-width-size relation in molecular clouds. 
Therefore, we adopt the central
hypothesis that {\em outflow-driven turbulence transports the 
flux of scalar momentum per unit mass (i.e, an acceleration) from the
driving scale to small scales}, where it is destroyed by the collision
of oppositely-directed motions.  The remainder of this paper amounts
to an exploration of this hypothesis.

Consider the one-dimensional velocity dispersion, or line-width,
$\sigma(r)$.  To be specific we define $3\sigma(r)^2/2$ to be the mean
kinetic energy per unit mass of spheres of radius $r$, relative to
their centers of mass.
%
Since for now we consider the gas
infinitely cold (an assumption we drop in \S\,\ref{Spectrum}),
$\sigma\to0$ as $r\to0$.  On small scales ($r\ll \rb$), $\sigma(r)$ is
determined by dimensional analysis:
\begin{equation} \label{sigma-r}
\sigma(r)^2 = \Bcoeff^2 {\Sb\Ib\over \rhob} r ~~~~(r\ll\rb)
\end{equation}
for some coupling coefficient $\Bcoeff$.  

A lower limit on $\Bcoeff$ can be estimated from the fact
that $\sigma(r)^2$ includes the kinetic energy of decelerating
shells.  The mean kinetic energy density in shells traveling faster
than $v$ is 
\begin{equation}
\Sb \int_0^{t(v)} {\Ib v'\over 2} d t' 
\end{equation}
where $t'(v')$ is the age at which the flow decelerates to $v'$.
Since the shell expands according to $dr' = v' dt'$, the integral can
be written
\begin{equation}
\Sb \int_0^{r(v)} {\Ib\over 2} dr' = {r(v)\over2}\Sb \Ib,  
\end{equation} 
which is $\Sb \Ib \rb/2$ at the merging scale, $r(v)=r(\sigma)=\rb$.
The total energy density on this scale is $3\sigma^2(\rb)/2$;
comparing to equation (\ref{sigma-r}), we find 
\begin{equation}\label{Bmin} 
\Bcoeff^2>{1/3}. 
\end{equation}
This is only approximate, since our estimate of the kinetic energy did
not account for the complicated dynamics of merging, and since
equation (\ref{sigma-r}) is expected to break down at the merging
scale.  However, it guides us to expect $\Bcoeff^2$ to be of order
unity, and to attribute a value in excess of $1/{3}$ to the
persistence of turbulent energy.  In fact, the
\cite{2006astro.ph..8471K} estimate of turbulent driving and decay
corresponds to a change in $\Bcoeff^2$ of 0.64 for this persistent
turbulence, so our best estimate of $\Bcoeff^2$ is $1/3+0.64 =0.97$.

On scales large enough that no individual outflow can dominate the
motion, the velocity spectrum is likely to be affected by a cascade
from even larger scales.  Indeed, if outflows are sufficiently weak or
the driving is sufficiently strong, it is possible for this cascade to
dominate over outflow-driven motions even on scales smaller than
$\rb$.   In the equations below we shall include a term to describe
this external cascade. 

\subsection{Variations of outflow intensity}\label{variations} 
Up to this point we have considered outflows of uniform intensity.
However fluctuations should be commonplace -- due to variations in
the mass and accretion rate of the driving star, for instance, or due
to the angle from which a collimated outflow is viewed.  

To treat this possibility, consider a single outflow whose momentum
scale $\IbOne$ differs from the characteristic value $\Ib$.  It 
dominates the motion of gas it overtakes within a ``merging'' radius
$\rmOne$, defined such that 
\begin{equation} \label{merging} 
M(\rmOne) \sigma(\rmOne)  = \IbOne. 
\end{equation} 
This equation relies on the basic assumption that the one-dimensional
velocity dispersion $\sigma(r)$ can be used to gauge the merging
radius, without reference to the intermittency of the motions on scale
$r$.  Moreover it neglects magnetic forces, which can cause merging at
smaller radii.  Here $M(\rmOne)$ refers to the mass encountered within
a radius $\rmOne$ of the driving source; we write
\begin{equation} \label{m-rho_r^3}
M(r) = {4\pi\over3} \phi_m \rhob r^3
\end{equation} 
where $\phi_m>1$ if stars tend to form within extended overdensities.
In general $\phi_m$ is a function of the other dimensionless
parameters, including $r/\rb$, but we shall treat it as a number. 

We previously considered the outflow strength $\Ib$ and rate $\Sb$ as
unique values.  We generalize now to a range of values for $\Ib$, and
consider $\Sb(<\Ib)$ as the cumulative rate of outflows weaker than 
$\Ib$.  To proceed we suppose that $\sigma(r)^2$ -- which is
proportional to the specific kinetic energy of motions on scales
smaller than $r$ -- is a linear superposition of the contributions
from outflows of a single intensity.

Therefore, let us reexamine the single-intensity limit
considered in \S\,\ref{oom}.  This is given by $\Sb(<\Ib)\rightarrow
S_1 H(\Ib-\Ib_1)$, where $H(x)$ is the Heaviside function, and where
subscript `1' denotes a single value of $\Ib$; then $d\Sb/d\Ib = S_1
\delta(\Ib-\Ib_1)$.  The contribution $\sigma_1^2(r)$ is $\propto r$ on
scales smaller than $\rmOne$, and is constant on larger scales;
therefore we adopt the idealization
\begin{equation} \label{sigmaSquaredOne}
\sigma_1^2(r) = \Bcoeff^2 {\Sb_1 \Ib_1\over \rhob} \left\{
\begin{array}{lc} r, & r<\rmOne \\ \rmOne, & r>\rmOne \end{array}. 
\right.
\end{equation} 
The contribution of outflows to $\sigma_1^2$ on scales larger than
$\rmOne$ should not be interpreted as an inverse cascade: it simply reflects
the fact that $\sigma^2(r)$ incorporates all motions within spheres of
size $r$.  Differentiating twice,
\begin{equation} 
{d^2\sigma_1^2\over dr^2} = - {\Bcoeff^2 \Sb_1\Ib_1 \over \rhob}
\delta(r-\rmOne)  =  - {\Bcoeff^2 \Ib_1 \over \rhob}
\frac{d\Sb/d\Ib}{d\rmOne/d\Ib}. 
\end{equation}
This equation holds for each value of $\Ib$ in our linear
superposition model, so in general 
\begin{eqnarray} \label{d2Sigma2dr2_formal} 
{d^2 \sigma^2\over dr^2} &=& -  {\Bcoeff^2 \Ib \over
  \rhob} {d\Sb\over dr}. 
\end{eqnarray} 

In this equation, as in the rest of this section, we take $\Ib$ to
mean the value for which $\rmOne = r$ according to equation
(\ref{merging}), so that $\Ib$, $\Sb$, and $r$ are monotonic functions
of one another; $d\Sb/dr$ could be replaced with
$(d\Sb/d\Ib)(d\Ib/dr)$.   Keep in mind that the relations
between these quantities depend on $\sigma(r)$, which must be
determined self-consistently. 

Integrating, 
\begin{equation}\label{dsigma^2/dr}
{d\sigma^2\over dr} = a_{\rm ext}  +
{\Bcoeff^2\over\rhob} \int_{\Sb(r)}^{\Stot} \Ib' d\Sb'
\end{equation} 
where $a_{\rm ext}$ is the acceleration due to a cascade, if any, from larger
scales, and $\Stot$ is the total outflow rate per unit volume.   Therefore 
\begin{equation} \label{dsigma^2dr_as_r->0}
\lim_{r\rightarrow0} {d\sigma^2\over dr} = a_{\rm ext} + {\Bcoeff^2\over\rhob}
\int_0^{\Stot} \Ib\, d\Sb, 
\end{equation} 
which generalizes equation (\ref{sigma-r}).  
Taking $\sigma^2$ to include thermal motions, we also have 
\begin{equation}\label{sigma^2_as_r->0}
\lim_{r\rightarrow0} \sigma^2 = c_s^2. 
\end{equation}

Given a total momentum injection rate $\int \Ib\, d\Sb$, what form for
$\Sb(<\Ib)$ maximizes $\sigma(r)$ on large scales?  It is obvious from
equation (\ref{SigmaAtRb}) that since $\rb/\tb\propto (\Ib
\Sb)^{3/7} \Ib^{1/7}$, turbulence is enhanced if momentum is emitted
in fewer, stronger bursts.  
(Up to a point: in a finite region, bursts that are too strong will
escape, and those that are too rare are irrelevant.)
Suppose we fix the {\em number} of
events and thus the mean outflow strength 
\begin{equation}\label{Ibbar}
\bar{\Ib} \equiv {\int \Ib\,d\Sb \over \Stot}
\end{equation} 
as well as $\int \Ib\,d\Sb$. Then, by the same argument, turbulence
is maximized if all but one outflow is negligible in strength, and the
last carries all the momentum.

More generally, we expect that turbulence is enhanced by broadening the
range of outflow strengths.  Consider the integration by parts of
equation (\ref{dsigma^2/dr}) over radius, from 0 to $r$:
\begin{eqnarray}\label{sigma(r)}
\sigma^2(r) &=& c_s^2 + r a_{\rm ext} +
       {\Bcoeff^2\over\rhob} r \int_{\Sb(r)}^{\Stot}  \Ib'\, d\Sb' 
\nonumber\\ &~& + 
 {\Bcoeff^2\over\rhob}\int_{0}^{\Sb(r)}  r' \Ib'\, d\Sb'. 
\end{eqnarray} 
This solves differential equation (\ref{d2Sigma2dr2_formal}), but only
implicitly, since the function $r(\Ib)$ given by equation
(\ref{merging}), and therefore $\Sb(r)$, depend on $\sigma(r)$.
Equation (\ref{sigma(r)}) is useful, however, because the first integral
disappears at the maximum outflow extent ($r = \max\rmOne$).  The
contribution of outflows to $\sigma^2(r)$ on even larger scales is
therefore given by the second integral
\begin{equation}\label{sigma2_on_large_scales}
 {\Bcoeff^2\over\rho}\int_{0}^{\Stot}  r \Ib\,  d\Sb. 
\end{equation} 
This allows us to define a {\em turbulence enhancement factor} as the
ratio of the outflow-driven turbulent pressure
(eq.~[\ref{sigma2_on_large_scales}]) to what it would be for a
monolithic model in which $\Ib=\bar\Ib$:
\begin{equation}\label{Enhc}
\Enhc \equiv 
{ \int_0^{\Stot} r \Ib d\Sb
 \over
  \left({3\over4\pi\phi_m}{\bar\Ib^4\Stot^3\over \Bcoeff\rhob^{1/2}}
    \right)^{2/7}}
\end{equation}
This factor illustrates the increase in $\sigma^2$ expected due to the
broadening of the outflow distribution.  It depends on $c_s$ and
$a_{\rm ext}$ because of the implicit nature of equation
(\ref{sigma(r)}), but to be definite we calculate it assuming $c_s =
a_{\rm ext} = 0$.  

Our expectation that broader distributions of
outflow strengths lead to stronger turbulent motions is verified in
Table  \ref{EnhcTable}, where $\Enhc$ is presented for the outflow
distributions discussed below. 

\subsubsection{Mass function} \label{IMF} 
We now consider the two primary sources of variation in outflow
intensity: differences in the stellar mass, in this subsection, and
differences with angle in collimated winds in the next subsection.

Let $F(>\Mstar/\Mbarstar)$ be the fraction of stars that exceed the
mean stellar mass $\Mbarstar$ at birth; then $F(0)= 1$ and
$\int x |dF/dx| dx= 1$ if the limits of integration include all stellar
masses.  Assuming $\Ib$ is strictly proportional to $\Mstar$ as
in equation (\ref{I=Mstar_vch}), and parametrizing the star formation
rate using equation (\ref{SFR}), we have
\begin{equation}\label{RateWithImf}
\Sb(>\Ib) = \left(32\over3\pi\right)^{1/2}\SFRff {G^{1/2}\rhob^{3/2}\over
\Mbarstar} F\left(>{\Ib\over\Mbarstar \vch}\right). 
\end{equation} 
The corresponding $\sigma(r)^2$ achieves a value enhanced by 
$\Enhc =1.85$ relative to a single-mass model.

\subsubsection{Outflow collimation}\label{collimation} 
A second and even more significant source of variation in outflow
intensity is the collimation of protostellar
winds. \cite{1999ApJ...526L.109M} argued that magnetic stresses cause
both disk winds and X-winds to approach the same asymptotic structure
at large distances from the source: the apparent strength at an angle
$\mu$ to the outflow axis is 
\begin{equation}\label{windstructure} 
\hat{\Ib}(\mu)  = \Ib \, P(|\mu|), ~~ P(|\mu|) = 
  \frac{1}{\ln\left({2/\theta_0}\right) 
  (1+ \theta_0^2 - \mu^2)}
\end{equation} 
where the normalization factor $\ln(2/\theta_0)$ ensures $\int_0^1
P(|\mu|)d|\mu| = 1$ so long as $\theta_0\ll 1$.  The softening angle
$\theta_0$ encompasses precession, internal shocks, fluid and magnetic
instabilities, phenomena at the jet-ambient interaction, and anything
else that degrades collimation.  \cite{1999ApJ...526L.109M} estimate
$\theta_0\sim 10^{-2}$ for several well-observed outflows.  This
implies a variation of $\sim 10^4$ in outflow strength with angle,
which has a strong effect on the amplitude and character of
outflow-driven turbulence.  (It is possible, however, that a larger
value of $\theta_0$ applies to the later stage of outflow evolution
and to the driving of turbulence.) 

Given an outflow of net momentum $\Ib$, the apparent strength equals
$\hat{\Ib}$ at angle 
\begin{equation} \label{PInverse}
|\mu| = P^{-1}(\hat{\Ib}/\Ib) = \left[1+\theta_0^2-{1\over
 \ln(2/\theta_0)\hat{\Ib}/\Ib}\right]^{1/2}, 
\end{equation} 
where $P^{-1}$ is the functional inverse of $P$, so long as this gives
$0\leq|\mu|\leq1$.  For a random orientation $|\mu|$ is evenly
distributed within this range; therefore $P^{-1}(\hat{\Ib}/\Ib)$ is
the cumulative probability that the apparent strength is weaker than
$\hat{\Ib}$, and $dP^{-1}/d(\hat{\Ib}/\Ib)$ is the distribution of
$\hat{\Ib}/\Ib$.  (Formally, $dP^{-1}/d(\hat{\Ib}/\Ib)=d|\mu|/dP$.)

We adopt the basic assumption that turbulence driven by collimated
outflows is very similar to turbulence driven by spherical outflows,
if one generalizes the distribution $\Sb(\Ib)$ to the {\em apparent
distribution} $\hat\Sb(\hat\Ib)$, which is the convolution of $\Sb$
and $P^{-1}$ over the logarithm of their arguments: 
\begin{equation}\label{Shat-Ihat}
{d\hat\Sb\over d\log \hat{\Ib}} = {d\Sb\over d\log \Ib}\otimes
    {dP^{-1}\over d\log \left({\hat\Ib/\Ib} \right)}. 
\end{equation} 
\begin{figure*}
\plotone{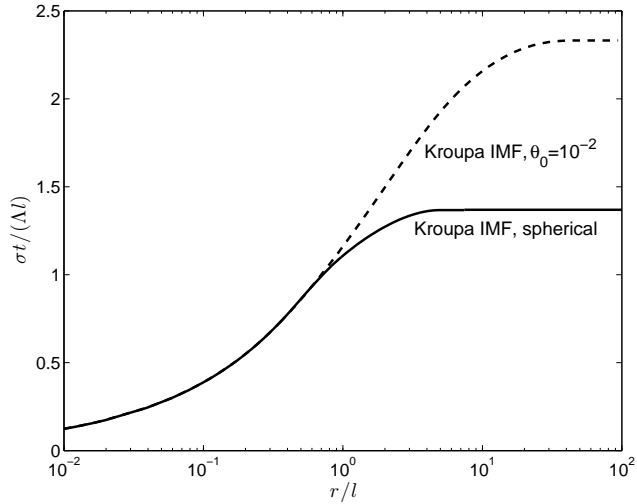}
\caption{Line-width $\sigma(r)$ versus size, normalized to the
  characteristic values in \S\,~\ref{oom}, for outflows driven by
  stars drawn from the \cite{2001MNRAS.322..231K} IMF.  If outflows
  are spherical, the theory of \S\,\ref{IMF} gives the lower curve; if
  they are collimated, the theory of \S\,\ref{collimation} gives the
  upper curve.  At scales much smaller than the driving scale $\rb$,
  $\sigma\propto r^{1/2}$ -- in agreement with the Larson's-law
  scalings for molecular clouds \citep{1981MNRAS.194..809L} as
  reported by \cite{1987ApJ...319..730S}. }
\label{r-sigma-nondim}\end{figure*}

\begin{deluxetable}{ccccc}
\tablecaption{Turbulence enhancement factors.}
\tablewidth{0pt}
\tablehead{
\colhead{Mass function} &
\colhead{$\Enhc$ (isotropic) } &
\colhead{$\Enhc$ ($\theta_0 = 10^{-2}$) }
}
\startdata
Single-mass &1.0 & 3.04 &  \\
 \cite{2001MNRAS.322..231K} IMF & 1.85 &   5.38
\enddata
{\label{EnhcTable}}
\end{deluxetable}
The influence of collimation on $\sigma(r)$ is illustrated in figure
\ref{r-sigma-nondim}, which displays numerical integrations of
equation (\ref{dsigma^2/dr}).  Two points can be drawn from these
results: (1) A finite thermal sound speed is of no practical
consequence so long as $\rb>c_s\tb$, since solutions converge when
$\sigma>c_s$; (2) $\sigma(r)$ continues to rise until $r\simeq
\max{(\rmOne)}$. (Recall that $\max(\rmOne)$ refers to the value
obtained by setting $\Ib\to\max\Ib$.)  However the rise in $\sigma(r)$
slows to zero as this limit is approached.

We expect the coupling coefficient $\Bcoeff$ to depend somewhat on the
degree of collimation, thanks to complicated phenomena like
collisions between young outflows \citep{2006ApJ...646.1059C},
although the sense and magnitude of this dependence are unknown. 

\subsection{Finite-size regions}\label{Nonlocality} 

What if the region of interest (a dense ``clump'', say) is too small
to catch all the outflow momentum?  The eruption of outflows from
protocluster regions is frequently observed, and this process deserves special
attention.  \cite{2000ApJ...545..364M} calculated the mass ejection
rate in this case; we wish to consider the outflows' dynamical
effects.

It is not sufficient to simply evaluate $\sigma(r)$ at the clump size
$R$ using the results of the previous section.  Equation
(\ref{dsigma^2/dr}) assumes a homogeneous background on scales larger
than $r$, as $\sigma(r)$ reflects the downward cascade driven by
outflows merging on larger scales, as well as the upward cascade
composed by those outflows as they expand.  However, outflows take
momentum as well as mass with them when they escape.  Collimation
allows this to happen in some directions without the entire clump
being unbound.  Specifically, escape occurs in direction $\mu$ when
$\hat\Ib(|\mu|) > \Iesc$, where
\begin{equation} \label{Iesc} 
\Iesc = c_g M \vesc
\end{equation} 
if $M= M(R)$ and $\vesc=(2GM/R)^{1/2}$ are the mass and escape
velocity of the region.  
The factor $c_g$ accounts for gravitational
deceleration of an expanding shell, which saps its momentum.  It is
close to unity, however: for density distributions $\rho(r)\propto
r^{-k_\rho}$, \cite{2000ApJ...545..364M} found 
\begin{equation} \label{cg}
c_g = \left(9-3k_\rho \over8-3k_\rho\right)^{1/2} 
\end{equation} 
(their equation [A13]).  This formula holds when outflows are driven
impulsively -- a safe assumption, so long as the wind that drives an
outflow has a duration similar to the free-fall time of its
collapsing, overdense core.

Given that outflow intensities exceeding $\Iesc$ get away, the theory
of \S\,\ref{variations} is easy to modify: simply replace $\Stot$
with $\Sb(\Iesc)$, throwing away the remaining momentum.  It is
important to realize that some outflows emerge from the clump surface
and rain back on it later; the replacement just suggested treats them
no differently from those that merge within the clump.  
Although approximate, this approach captures the essential division
between capture and escape. 

\begin{figure*}
\plotone{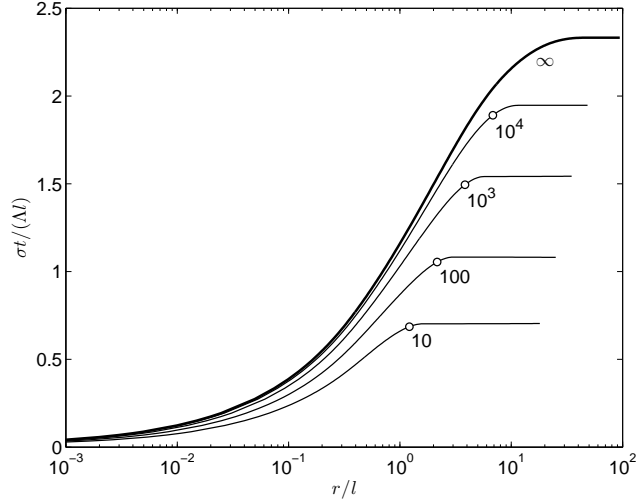}
\caption{Effect of finite clump radius on turbulence driven by
  collimated outflows ($\theta_0 = 10^{-2}$) driven by stars drawn
  from the \cite{2001MNRAS.322..231K} IMF.  The thick curve represents
  an infinite uniform medium; the other curves are labeled by
  $\Iesc/\bar\Ib$, which determines how much momentum is lost from a
  finite clump.  Circles mark the edge of the region for fiducial
  parameters $n_{H4}=1$, $\vch=40$\,km\,s$^{-1}$, $\SFRff =0.034$, and
  $\Bcoeff=1$. }
\label{escaping}\end{figure*}

The effect of outflow eruptions on the velocity scale $\sigma(R)$ is
determined by how much outflow momentum is eliminated in this
procedure.  It depends, therefore, on the dimensionless ratio
${\Iesc/\bar\Ib}$ as well as on the shape of $\Sb(\Ib)$ (see also
\S\,\ref{Grav}).  This dependence is illustrated in figure
\ref{escaping}, 
where a slow dependence on $\Iesc/\bar\Ib$ is
apparent.  This is natural, because the apparent intensity $\hat\Ib$
can exceed $\bar\Ib$ by a factor of $10^{6.7}$ in the model plotted --
of which $10^{2.7}$ comes from the range of stellar masses, and $10^4$
arises from collimation.  
For a default model in which we adopt the \cite{2001MNRAS.322..231K}
IMF, take $\vch$ to be independent of $\Mstar$, consider winds to
be collimated with $\theta_0=10^{-2}$, and take $a_{\rm ext}=0$, we find 
\begin{equation} \label{sigma_Model}
{\sigma^2 \tb^2 \over \rb^2 } \simeq {c_s^2 \tb^2 \over\rb^2} + 
\Bcoeff^2 \left[K_1^{-2\eta} + K_2^{-2\eta}
  \left(r\over\rb\right)^{-\eta}\right]^{-1/\eta} 
\end{equation} 
where the fit parameters $K_1$, $K_2$, and $\eta$ are functions of
${\Iesc/\bar\Ib}$: 
\begin{equation} \label{K_1_Model}
K_1 = \left[2.33^{-10} +
  \left(0.32+0.40\log_{10}{\Iesc\over\bar\Ib}\right)^{-10}\right]^{-1/10}, 
\end{equation} 
\begin{equation} \label{K_2_Model} 
K_2 = \left[1.22^{-6} +
  \left(0.55+0.26\log_{10}{\Iesc\over\bar\Ib}\right)^{-6}\right]^{-1/6}, 
\end{equation} 
and 
\begin{equation} \label{eta_Model} 
\eta ={11 + 15 \log_{10}(\Iesc/\bar\Ib) \over 
1 + 10 \log_{10}(\Iesc/\bar\Ib) }. 
\end{equation} 
This fit reproduces the numerical evaluation of $\sigma(r)$ to within
5\% for $\Iesc> 6 \Ib$.  The quantities used in these
formulae are, in convenient form,
\begin{equation}\label{IescOnIb_eval}
{\Iesc\over \bar\Ib } = 770\, M_3^{5/4} \Sigma_{\rm cgs}^{1/4} {40\,\rm
  km\,s^{-1}\over\vch}, 
\end{equation} 
\begin{equation}\label{rb_eval}
\rb = 0.079\, M_3^{5/28}  \Sigma_{\rm cgs}^{-15/28} \left({40\,\rm
  km\,s^{-1}\over\vch}{0.034\over\SFRff }\right)^{1/7}\,\rm pc, 
\end{equation}
and 
\begin{equation}\label{rb/tb_eval}
{\rb\over\tb} = 1.24\, M_3^{-1/28}  \Sigma_{\rm cgs}^{3/28}
  \left({\vch\over 40\,\rm km\,s^{-1}}\right)^{4/7}
 \left({0.034\over\SFRff }\right)^{3/7}\,
  {\rm km\over s}
\end{equation}
where $M(R) = 10^3 M_3 \,M_\odot$ and $M(R)/(\pi R^2) = \Sigma_{\rm
  cgs}$\,g\,cm$^{-2}$, and we used $k_\rho=1.5$ to evaluate $c_g$. 

\section{Comparison to simulation} \label{sim} 

The simulation by \cite{2006ApJ...640L.187L} offers the closest
comparison to the theory presented above.  These authors simulate a
self-gravitating, weakly magnetized, initially turbulent cloud,
identifying regions of localized collapse and replacing them with
spherical outflows.  Using their fiducial conversion from code units
to physical units, their outflow momentum scale is nearly constant at
$\Ib = 10^{39.7}\rm g\,cm\,s^{-1}$.  The 960 $M_\odot$ of cloud gas is
initially arranged in a mildly centrally peaked density profile of outer
radius 0.75 pc, and becomes more centrally condensed over the course
of the run.  After 0.6 Myr star formation begins in earnest, and 
between 0.9 and 1.2 Myr the stellar fraction increases from 6.5\% to
13.4\% of the cloud mass.   \citeauthor{2006ApJ...640L.187L} report a
final, three-dimensional rms velocity of 1.5\,km\,s$^{-1}$
corresponding to $\sigma=0.85\rm\,km\,s^{-1}$. 

If one assumes that the gas has contracted by a fraction $f_R$ of
its initial radius during the period 0.9-1.2 Myr, then $\rhob \simeq
10^{-19.4}f_R^{-3}\,\rm g\,cm^{-3}$ and $\Sb \simeq
10^{-66.6}f_R^{-3}\,\rm cm^{-3}\,s^{-1}$ in that period.  Equation
(\ref{sigmaSquaredOne}) then predicts $\sigma = 1.7
f_R\Bcoeff\rm\,km\,s^{-1}$, which is consistent with the
\citeauthor{2006ApJ...640L.187L} results if $\Bcoeff = 0.6
f_R^{-3/7}$.  Contraction by $f_R=1/2$ appears reasonable based on
their figure 2, in which case \[\Bcoeff \simeq0.8.\] This seems
entirely consistent with our expectations from \S\,\ref{oom} that
$\Bcoeff\sim 1$.  Note, however, that one should expect changes in
numerical resolution, cloud magnetization, and outflow collimation to
be accompanied by variations in $\Bcoeff$.   

\citeauthor{2006ApJ...640L.187L} also mention that collimated outflows
appear to drive more vigorous turbulence, as we would expect from
\S\,\ref{collimation}.  \cite{LiNakamura06} have modified their
previous simulations by collimating outflows within 30$^\circ$ half
opening angle, and report that the enhancement is not discernible from
the simulations.  In our theory, this modification amounts to
increasing $\Ib$ by a factor of 7.5 while holding $\Sb\Ib$ fixed; we
would expect $\sigma$ to increase by $7.5^{1/7} = 1.3$ (as in eq.\
[\ref{SigmaAtRb}]), so long as $\Bcoeff$ does not change.  Whether
this mild enhancement of turbulent velocity is seen in simulation will
require further investigation. 
In an equilibrium cloud, as our referee points out, one must tease
this effect from the redistribution of mass caused by more vigorous
turbulence.

A deeper investigation of the same class of model is reported by
\cite{NakLi07}, who report several results in line with our
predictions.  In this work, outflows posses a 30$^\circ$ ``jet''
component and a uniform ``wind'' component.  The authors report that
the jet component is more effective at driving turbulence, based on
the slowdown of star formation observed if jets are strengthened (all
else being equal).  Their turbulent power spectrum obeys $E(k)\propto
k^{-2}$ as anticipated in equation (\ref{spectrum}), and flattens at
large scales -- possibly reflecting the flattening of $\sigma(r)$
(figure \ref{r-sigma-nondim}).  Finally, they obtain an equilibrium
density profile $\rho\propto r^{-1.5}$, in line with the argument we
present below in \S \ref{implic}.
   
\cite{MacLow99outflows} has also simulated turbulence driven by collimated
sources, but a direct comparison is not possible because the
driving field in his simulation was steady rather than impulsive. 

\section{Comparison to observation: NGC 1333} \label{obs}

Located within the Perseus molecular cloud, the NGC 1333 reflection
nebula is the site of a vigorous burst of clustered star formation
within a dense molecular clump.  The current population of at least
143 young stars is estimated
to be 1-2 Myr old \citep{1996AJ....111.1964L}, and about ten molecular
outflows and Herbig-Haro systems emerge from the most recently formed
objects. \cite{2003ARA&A..41...57L} estimate a stellar mass of
$\sim79\,M_\odot$, based on the limiting $K$ magnitude of $14.5$.  
Distance estimates range from $D=$210~pc to 350~pc, as discussed by
\cite{1996A&A...306..935W}. 

\cite{2003AJ....126..286R} map the NGC 1333 core (and many others) in
$^{13}$CO(1-0), C$^{18}$O(1-0), and C$^{18}$O(2-1), probing
sequentially smaller and denser regions.  They derive masses from a
large velocity gradient analysis; combining these with their reported
velocity dispersions, and assuming that gas dominates the mass budget,
we derive the virial parameter \citep{BM92}
\begin{equation}\label{alpha1333}
\alpha_{\rm vir} \equiv {5R\sigma(R)^2 \over G M(R)} = \left(1.27,
     1.32, 1.19\right) { 220\,\rm pc\over D}
\end{equation}
for $^{13}$CO(1-0), C$^{18}$O(1-0), and C$^{18}$O(2-1), respectively.
Since $\alpha_{\rm vir}$ is thought to vary, in strongly
self-gravitating regions, between 1.11 \citep[for molecular
clouds;][]{1987ApJ...319..730S} and 1.34 \citep[for massive,
magnetized molecular cores;][]{2003ApJ...585..850M}, we adopt $D=220$
pc \citep[close to $D=212$ pc adopted by][]{2005ApJ...632..941Q}.  The
embedded star cluster then have an effective radius $R=0.39$\,pc, scaling
the value given by \cite{2003ARA&A..41...57L} to the closer distance;
the enclosed mass is then 446\,$M_\odot$, and the escape velocity is
3.1\,km\,s$^{-1}$.  The velocity dispersion at the same radius is
$\sigma=1.1\,\rm km\,s^{-1}$.

Can this velocity dispersion be maintained by the protocluster
outflows?  To evaluate the formulae in \S\,\ref{variations} and \S
\ref{Nonlocality}, we adopt the \cite{2001MNRAS.322..231K} IMF and
assume the current star formation rate equals 79\,$M_\odot$ per Myr.
Other model parameters are $\Stot = 10^{-65.8}\rm
cm^{-3}\,s^{-1}$, $\rho_0 = 10^{-18.9}\rm g\,cm^{-3}$, $T=20$ K,
$\theta_0=10^{-2}$, and $\Iesc = 10^{41.5}$\,g\,cm\,s$^{-1}$ (where we
have used $k_\rho=1.35$ to estimate $c_g=1.1$ in eq.~[\ref{cg}]).  For
reference, these values give $\SFRff\simeq 0.034$ in the region of
interest -- virtually identical to the value $\SFRff=0.038$ we derive
from the \cite{2005ApJ...630..250K} theory.  We take $a_{\rm ext}=0$
to probe the influence of outflows alone.

Using $\vch = 40$\,km\,s$^{-1}$ and applying the prescription of \S
\ref{Nonlocality} to predict the velocity dispersion at $R=0.39$\,pc,
we find $\sigma(R)=1.12 (\Bcoeff/0.8)$\,km\,s$^{-1}$.  This is
entirely consistent with the observed value of 1.1\,km\,s$^{-1}$, for
precisely the same value of $\Bcoeff$ we estimated in \S\,\ref{sim}.
This result is entirely consistent with the proposition that
turbulence in the star-forming region of NGC~1333 has been regenerated
by outflows. The cluster outflow inferred by
\cite{1996A&A...306..935W} could then legitimately be viewed as the
mass ejected by erupting jets, and the inflow detected by
\cite{2006ApJ...637..860W} might represent a return flow of gas
ejected below the escape speed.

Although the value $\vch = 40$\,km\,s$^{-1}$ (which was justified in
\S~\ref{vch} using the observational relations reported by
\citealt{Richeretal}) is in line with the momentum of the NGC~1333
outflow HH~7-11 as estimated by \cite{1981ApJ...251..103S},
\cite{2000A&A...361..671K} and \cite{2005ApJ...632..941Q} have
inferred significantly lower momenta for the NGC~1333 outflows.
\citeauthor{2000A&A...361..671K} estimate $\sim 1
M_\odot$\,km\,s$^{-1}$ per outflow for ten currently active outflows,
and \citeauthor{2005ApJ...632..941Q} derive a similar momentum scale
for the explosions that open cavities in the gas.  Given that the mean
stellar mass in the \cite{2001MNRAS.322..231K} IMF is $0.21\,M_\odot$,
this momentum scale corresponds to $v_c\sim 5$\,km\,s$^{-1}$.  In the
formalism of \S\,\ref{Nonlocality}, this lower value would imply
$\sigma(R) = 0.47 \Bcoeff$\,km\,s$^{-1}$.

It is possible that, for strongly collimated outflows in a real,
magnetized star-forming region like NGC\,1333, $\Bcoeff=2.3$ as
required to make this lower estimate consistent with the observed
value of 1.1\,km\,s$^{-1}$.  However it is equally probable that the
inference $\vch\sim 5$\,km\,s$^{-1}$ is an underestimate: compare this
to the estimates in \S\,\ref{vch}, and note that the {\em minimum}
value estimated by \citeauthor{Richeretal} (\citeyear{Richeretal}, in
their \S\,III.B) is $\sim$30\,km\,s$^{-1}$.  One solution to this
discrepancy could be that the outflows now observed in NGC\,1333 are
simply smaller than the average -- possible, since the median value of
$M_\star$ in the \cite{2001MNRAS.322..231K} IMF is 2.6 times lower
than the mean value.  Alternatively these outflows could be more
powerful than \cite{2000A&A...361..671K} report.  The fact that they
find a $2.5\,M_\odot$\,km\,s$^{-1}$ as the net momentum of HH\,7-11 in
their $^{12}$CO(3-2) and $^{12}$CO(2-1) survey, whereas
\cite{1981ApJ...251..103S} find $\sim 34$\,km\,s$^{-1}$ for the same
outflow in observations involving the four transitions
$^{12,13}$CO(J=2-1,1-0), raises this possibility.  CO optical depth,
uncorrected inclination, motions at velocities close to systemic, and
loss of momentum from the cloud can all cause outflow momentum to be
underestimated.  (See \citealt{2005AJ....129.2308W} for a discussion
of some problems in the determination of outflow momentum.)  Moreover,
the most noticeable outflows are the young, rapidly expanding ones.
Since these are still being driven, they will not have acquired their
final momenta.  Finally, note that very little mass can be ejected
from the region by outflows if $\vch$ is as low as 5\,km\,s$^{-1}$:
the theory of \cite{2000ApJ...545..364M} predicts a star formation
efficiency $\varepsilon\simeq 88\%$ in this case, whereas
$\varepsilon\simeq 47\%$ if $\vch=40$\,km\,s$^{-1}$.  The latter value
is more consistent with the fact that only $\sim 18\%$ of the mass is
currently in stars.

For all of these reasons we consider it most likely that turbulence in
NGC\,1333 is driven by outflows with $\vch\sim40$\,km\,s$^{-1}$ (such
that $\Bcoeff \simeq 0.8$ in the theory of \S\,~\ref{Nonlocality}),
although the alternatives -- that $\Bcoeff \simeq 2.3$, or that
outflows are only a minor contributor to the turbulence -- cannot be
ruled out.  The last option is especially unattractive, as it leaves
unanswered the question of how turbulent energy is regenerated there.

\section{Implications for star cluster formation} \label{implic} 

Our analysis of NGC~1333 supports the assertion that outflow-driven
turbulence affects the dynamical state of gas clumps while star
clusters form within them.  How, then, does feedback influence the
creation of star clusters?

\begin{figure*}
\centerline{\includegraphics[width=0.8\hsize]
    {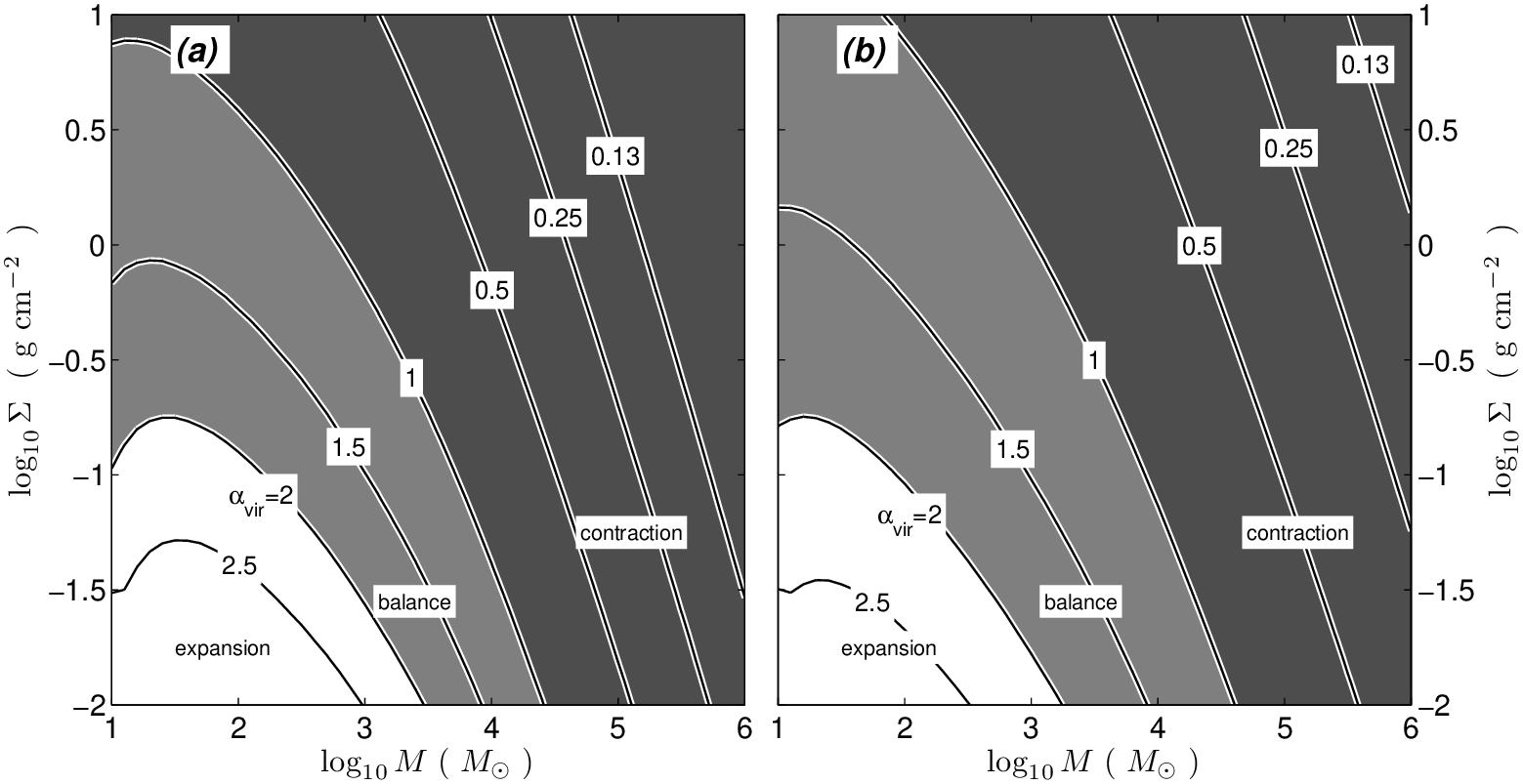}}  
\caption{Virial parameter $\alpha_{\rm vir}$ for outflow-driven
   turbulence within clouds of various total mass $M$ and column
   density $\Sigma=M/(\pi R^2)$.  Outflows are assumed to be
   collimated ($\theta_0=10^{-2}$) and driven, with
   $\vch=40$\,km\,s$^{-1}$, by stars drawn from the
   \cite{2001MNRAS.322..231K} IMF; $\Bcoeff=1$.  In {\em (a)},
   $\SFRff$ is held constant at 0.034; in {\em (b)}, $\SFRff$ is
   derived self-consistently from $\alpha_{\rm vir}$ using the star
   formation rate model of \cite{2005ApJ...630..250K}.  In both cases
   the gas temperature is assumed to be 20\,K throughout, and there is
   no external driving ($a_{\rm ext} = 0$).  Conditions for expansion
   and contraction are estimated assuming that virial balance
   corresponds to a value of $\alpha_{\rm vir}$ between 1 and 2. }
\label{alpha}\end{figure*}

A preliminary answer can be gleaned from figure \ref{alpha}, where we
calculate $\alpha_{\rm vir}$ for cluster-forming regions of various
total mass ($M$) and column density ($\Sigma = M/(\pi R^2)$) using the
formalism of \S\,\ref{Nonlocality}.  In this calculation the region was
assumed to be neither expanding nor contracting, and devoid of any
additional source of turbulence ($a_{\rm ext}=0$); outflows were
collimated with $\theta_0 = 10^{-2}$ and $\vch=40$\,km\,s$^{-1}$, and
the \cite{2001MNRAS.322..231K} IMF was adopted.  In the left panel,
the star formation rate parameter $\SFRff$ was held fixed at the value
0.034 we derived for NGC~1333; on the right, $\SFRff$ and $\alpha_{\rm
vir}$ were derived self-consistently according to the model of
\cite{2005ApJ...630..250K}; however the gas temperature was held fixed
at 20\,K for simplicity.

We expect that there exists a critical value of $\alpha_{\rm vir}$, 
between 1 and 2, such that a cluster-forming clump will be in
virial equilibrium given a moderate external pressure; for instance, 
\cite{2003ApJ...585..850M} derive  $\alpha_{\rm
  vir}=1.34$ for their equilibrium model of magnetized, turbulent cores.
Smaller values of $\alpha_{\rm vir}$ imply contraction.  Higher values
imply either expansion, or require a strong confining pressure. 

An immediate conclusion from figure \ref{alpha} is the existence of an
outflow-driven equilibrium state ($\alpha_{\rm vir}\simeq 1.5$) for
column densities and masses relevant to star cluster formation.  It is
however an {\em unstable} equilibrium: since $\partial \alpha_{\rm
vir}/\partial \Sigma<0$ at fixed mass, turbulent support is weakened
if an equilibrium clump is compressed.  The instability is reduced --
though not removed -- by the self-regulation of the star formation
rate in the \cite{2005ApJ...630..250K} model.  It would be reduced
even further if we were to account for heating of the gas by star
formation, as the sound speed enters (weakly) into the
\citeauthor{2005ApJ...630..250K} theory.  

Unstable equilibria were previously found by \cite{M89} in models for
giant molecular clouds, and by \cite{myphd} and
\cite{1999sf99.proc..353M} in simpler models for star cluster
formation. In fact, instability is generic to any model in which
$\SFRff$ and $\vch$ are slowly varying functions of the cloud
parameters.  For a virialized cloud this can be seen by comparing the
cloud dissipation rate $\sim \sigma(R)^5/G$ with its energy input
rate -- which scales as the star formation rate $\sim \SFRff
\sigma(R)^3/G$, times the energy injection per unit star mass $\sim
\vch \sigma(R)$.  So long as $\SFRff\vch$ does not increase as fast as
$\sigma^2$ when the cloud contracts, dissipation overcomes energy
injection.  Sharp thresholds, such as the photoionization column in
\cite{M89}, are therefore required for stable equilibria. 

The scenario for star cluster formation implied by figure \ref{alpha}
is qualitatively consistent with the one advanced by 
\cite{2006ApJ...641L.121T} and \cite{2006astro.ph..6277K}, in that
turbulence supplied by outflows is important enough to slow collapse
over several, perhaps many, crossing times.  Because the equilibrium
state is unstable, is also consistent with
proposal by \cite{2006ApJ...640L.187L} that this support may fail in
some circumstances, and that this might lead to a state of collapse
that triggers massive star formation.  However, massive-star outflows
are almost certainly more powerful than our simple model has assumed
-- as discussed in \S~\ref{vch} and by \cite{2002ASPC..267..267T}.
If there is a column density threshold for massive star formation,
then it is possible that the feedback from these stars creates a
stable equilibrium -- in which $\Sigma$ remains near the threshold.
Alternatively, massive stars might disrupt the clump outright; we
leave these questions for future investigation.

The form of the line-width-size relation $\sigma(r)$ affects the
internal structure and dynamics of a clump, in addition to its global
expansion or contraction.  Suppose we define $\beta(r)=d\ln\sigma/d\ln
r$ as the logarithmic slope of this relation.  In a singular
polytropic model for the clump \citep{2003ApJ...585..850M}, the
density index $k_\rho$ (in $\rho\propto r^{-k_\rho}$) is related to
$\beta$ through $k_\rho=2-2\beta$ 
(and both $\beta$ and $k_\rho$ are assumed constant in radius).
In our model, $\beta(R)$ can take any value between 0 and 1/2,
depending on the ratio between $R$ and the turnover in $\sigma(r)$ --
as seen in figure \ref{escaping}.  This figure also shows that
star-forming regions with $n_H\sim 10^4$\,cm$^{-3}$ have radii near
this turnover, so that $\beta(R)\simeq 1/4$ -- generally
consistent with the value $\beta = 0.21\pm0.03$ obtained for massive
cores and clumps by \cite{1995ApJ...446..665C}.  More specifically,
the fit given in equation (\ref{sigma_Model}) implies
\begin{equation} \label{beta_Model} 
\beta\simeq \frac{1}{2H \left( 1 + {c_s^2 \tb^2 \over \Bcoeff^2 R \rb}
H\right)} 
\end{equation}
where 
\begin{equation}
H = 1 + \left(\frac{K_2}{K_1}\right)^{2\eta} 
\left(\frac{R}{\rb}\right)^{\eta}. 
\end{equation} 
To the extent that the \cite{2003ApJ...585..850M}
model can be generalized to our model for turbulence, $\beta\simeq
1/4$ implies
$k_\rho\simeq 1.5$ on the scale of the region.  This estimate
coincides with the value \cite{2003ApJ...585..850M} adopted after a
survey of the observational literature.  If our proposal for the
origin of $k_\rho$ is correct, then one would expect that $k_\rho$ is
steeper (closer to 2) in denser and more massive regions (so long as
turbulence is dominated by outflows), as both $R/\rb$ is higher there,
and as $\Ib/\Iesc$ is lower.  Note that \cite{2006ApJ...644..355H}
derive $k_\rho\simeq2$ for the birth region of the Orion Nebula
Cluster.

The line-width-size index $\beta$ also enters into estimates for the
typical radii of protostellar disks.  
\cite{2006astro.ph..9692K} calculate the specific angular momenta of
cores prior to their collapse, and find that a larger value of $\beta$
leads to a greater angular momentum.  Note that, since $\beta$ is a
decreasing function of $r$, its value is greater on the core scale
than on the clump scale.
\citet{2006ApJ...641L.121T} have previously proposed that $\beta\simeq
1/2$ on small scales, falling to $\sim 1/4$ on scales large
enough to sample the tidal field.  We propose that this shift is
produced by the dynamics of driving -- and shapes the gravitational
potential, rather than reflecting it.

\section{Caveats}\label{caveats}

The theory presented here relies on several assumptions and
approximations.  The central hypothesis, as stated in \S
\ref{Spectrum}, is that a strongly dissipative, supersonic turbulence
can be described as a cascade of (scalar) momentum to small scales in
much the same way that Kolmogorov turbulence can be described as an
energy cascade.  This is not too bold, as it agrees with the spectral
slopes observed in simulations of these cascades.  In
\S\,\ref{variations} we assumed that these cascades will simply add, in
the case that turbulence is driven by more than one type of source.
In \S\,\ref{collimation} we assumed, further, that this 
superposition can be applied to collimated outflows -- if the
variation of strength with angle is treated in the same manner as a
population of isotropic sources of different strengths.  Moreover, our
treatment in \S\,\ref{Nonlocality} assumes that the turbulent velocity
in a region of finite radius can be estimated by (1)
discarding momentum ejected in escaping outflows and (2) evaluating
the resulting line-width-size relation at the scale of the region.    
Finally, we have assumed that the differences between outflow-driven
turbulence in magnetized and unmagnetized gas can be encompassed by
modifying the coupling coefficient $\Bcoeff$, and in \S\,\ref{implic},
by a modifying the critical virial parameter corresponding to
collapse.
All of these assumptions are sure to be approximate at some level, and
must be tested against numerical simulations and observations more 
thoroughly than we have done in \S\,\ref{sim} and \S\,\ref{obs}.

\section{Summary and Conclusions} \label{conclusions}

Based on our observation that the spectral slope of supersonic
turbulence implies a momentum cascade rather than an energy cascade,
we have constructed a simple model for turbulence stirred by momentum
injection from stellar outflows.  A key feature of our model is that
turbulence is enhanced if momentum injection is spread amongst
outflows of a wide range of strengths, or, by extension, if outflows
are strongly collimated.  These effects allow for some of
the momentum to be deposited on relatively large scales where
turbulent decay is slow.  The diversity of outflow strengths
(reflecting the range of stellar masses), and their strong collimation
by magnetic fields, both imply that real outflows are quite effective
at driving turbulence.  

Our comparison to NGC\,1333 supports the assertion that turbulence in
the cluster-forming gas has been entirely regenerated by outflows.
Although uncertainties regarding the outflow momentum scale and the
value of the theoretical outflow-turbulence coupling coefficient
$\Bcoeff$ limit the strength of this conclusion, these uncertainties
will be eliminated by future numerical and observational studies.

In our fiducial model, outflows will maintain turbulence within 
cluster-forming regions of a few thousand solar masses, whose column
densities are $\sim 0.3-1$\,g\,cm$^{-2}$.  Intriguingly, the energetic
equilibrium in these models is an unstable one.  It is possible that
this leads to global contraction or collapse, as
\cite{2006ApJ...640L.187L} have suggested.  However this instability
could equally be an artifact of our assumption that outflow
momentum scales in proportion to the mass of the forming star.

The structure of a turbulence-supported region reflects its turbulent
spectral slope.  The radial density index takes values close to -2 if
turbulence is driven on small scales, or close to -1 if it is driven
on scales larger than the region of interest.  Our model predicts
intermediate slopes ($\sim -1.5$) because outflow collimation allows
both of these to occur simultaneously.  We expect this slope, which is
consistent with observations of cluster-forming regions, to steepen in
regions that catch outflow momentum more effectively. 

Although our model was developed to address protostellar outflows, it
could in principle be extended to other forms of driven, supersonic
turbulence in radiative gas, such as HII-region-driven turbulence in
molecular clouds \citep{2002ApJ...566..302M} or supernova-driven
turbulence in the diffuse interstellar medium
\citep{2001astro.ph..6509M}.  


\acknowledgments 
This work was supported by an NSERC discovery grant and by the Canada
Research Chairs program.  The author is pleased to thank Kaitlin
Kratter, Chris McKee, and the referee for comments,emu and Zhi-Yun Li and
Fumitaka Nakamura for sharing some results prior to publication.





\end{document}